# Indexing Research Papers in Open Access Databases

**Alexandra-Emilia Fortiş**
"Tibiscus" University, Timişoara, Romania
afortis@tibiscus.ro

**ABSTRACT**. This paper synthesizes the actions performed in order to transform a classic scientific research journal – "Annals. Computer Science Series" – available only in printed form until 2008, into a modern e-journal with free access to the full text of the articles. For achieving this goal, the research papers have been included in various article databases, portals and library catalogs which offered a high visibility to the journal.

**Introduction**

Starting from the requirements of the National Center for Science Policy and Scientometrics (CENAPOSS - Romanian acronym for "*Centrul Naţional pentru Politica Ştiinţei şi Scientometrie*") with respect to the methodology of assessing scientific research journals, some of the criteria that have to be considered by the Editorial Board of a scientific research journal are:

- The inclusion of the journal in international catalogs;
- The inclusion of articles in international articles and abstracts databases.

The action initiated at Computers and Applied Computer Science Faculty within "Tibiscus" University of Timişoara focused on the identification of important articles and abstract international databases for the field of Computer Science and its applications. After the identification and the inspection of available databases, the Editorial Board of "Annals.

Computer Science Series" decided that the journal will be included in Open-Access databases.

Some of the advantages arising from the use of Open-Access databases are:

- A high visibility of the journal;
- Absence of inclusion costs;
- Absence of subscription fees;
- Readers do not need to register in order to view the papers or the abstracts;
- Increased possibility to share knowledge;
- Increased possibility of citation of papers, by other researchers.

## 1 Open Access Initiative. Open Archive Initiative

The Budapest Open Access Initiative was developed in 2001 by the Open Society Institute in Budapest, with the main goal of making research papers from the scientific world freely available via Internet.

Briefly, the Open Access Initiative intends to distribute, without any charges, peer-reviewed scholarly journals, with the dissemination of knowledge as only purpose. Also, this initiative offers a great advantage to researchers, offering them the possibility to publish their works without costs, through the on-line publishing of the papers. That was the impulse that transformed many scientific journals into modern, visible, easily readable and searchable e-journals. The concept was rapidly adopted by researchers, teachers and many other categories, appearing the wish to publish lecture notes, course materials, preprints or works in progress. With the rapidly increasing amount of research papers, a new challenge came up: the need of storage of many research papers. Classical databases weren't enough for such an action because older deposits had to be easily reachable, through simple searches in browsers. It is the moment when article databases (or archives) and repositories appear and, more, they had to be in accord with the frame established by the Open Access Initiative. The appropriate concept to handle this aspect was open archives.

In order to offer an open access character to scholarly journal, the authors of this initiative were proposing two complementary strategies: self-archiving and the creation of open-access journals.

Through open archives, one is enabled to use "*Web-accessible material through interoperable repositories for metadata sharing,*

132

*publishing and archiving. It arose out of the e-print community, where a growing need for a low-barrier interoperability solution to access across fairly heterogeneous repositories leads to the establishment of the Open Archives Initiative (OAI).*

*The OAI develops and promotes a low-barrier interoperability framework and associated standards, originally to enhance access to e-print archives, but now taking into account access to other digital materials*" [OAI02]. Also, the OAI mission states that "the Open Archives Initiative develops and promotes interoperability standards that aim to facilitate the efficient dissemination of content."

The OAI is an initiative to develop and promote interoperability standards that aim to facilitate the efficient dissemination of the content.

## 2 Metadata and Standards

Following the definition, metadata represents "*data that describes other data*". For research papers, metadata means information on titles, authors and their affiliation, contact data, the status of the work, the language used, the number of figures and tables in the paper, the number of pages, the abstract and keywords.

By using metadata, one can obtain additional information on the deposit. In the case of a journal, this can describe the conference (name and location) where the papers were presented. From another point of view, the metadata we've used content metadata which are immutable (the data considered do not change in time or as the article is read). Also, the metadata associated with a scientific paper can be characterized as active, meaning that it can be consulted by a generic computer program.

All digital libraries frequently use a set of information called metadata, in order to describe their objects (research papers, lecture notes, books, etc). This information can be classified as follows:

- Descriptive – used in relation with bibliographies and subjected to standards such as MARC (Machine-Readable Cataloging);
- Structural – with relations between different parts of the deposit (abstract, text, figures, tables) and with linkage information between deposits (e.g. articles in a proceedings volume);
- Administrative – related to the deposit storage format or its original format, aspects on license and copyright.

In the case of a scientific journal, metadata is most often used when cataloguing and searching.

In order to build accurate and correct metadata, one needs to implement from the beginning some standards for metadata in digital libraries, such as Dublin Core, METS, PREMIS schema and OAI-PMH.

The Dublin Core standard *"is a standard for cross-domain information resource description"*. It includes 15 elements, as follows: title, author, subject, description, publisher, contributor, date, type, format, identifier, source, language, relation, coverage and rights. Extracted directly from their mission, we can resume this initiative as a modality to easily find, share and manage information. [Dub09]

METS comes from Metadata Encoding & Transmission Standard and it is defined as *"a standard for encoding descriptive, administrative, and structural metadata regarding objects within a digital library, expressed using the XML schema language of the World Wide Web Consortium."* It was developed by the Library of Congress in Washington, USA. It is now used in major libraries around the world. [Met09]

PREMIS - **Pre**servation **M**etadata: **I**mplementation **S**trategies is a metadata scheme used together with METS, as an administrative tool. [Pre05]

OAI-PMH - Open Archives Initiative Protocol for Metadata Harvesting represents *"is a low-barrier mechanism for repository interoperability"*. Through OAI-PMH we have a technical tool able to harvest entries containing metadata in any format. It is based on other open standards such as HTTP and XML. It doesn't support search and retrieval activities but in conjunction with other instruments can solve those issues. [LS08]

## 3 Results

After research and analysis performed by the Editorial Board the journal "Annals. Computer Science Series" is now included in several scientific databases, abstract databases and portals dedicated to research as we shall briefly present. Two different databases were taken into account: arXiv and DOAJ.

ArXiv is developed, owned, operated and funded by Cornell University Library in New York, USA. We are dealing with an e-Print archive sustaining, together with ACM (the Association for Computing Machinery), NCSTRL (Networked Computer Science Technical Reference Library) and AAAI (Association for the Advancement of Artificial Intelligence), an instrument called Computing Research Repository (CoRR).

This instrument allows its users to search, browse and download paper, without charges.

Journal owners (editors, publishers) and article owners (authors or editors of volumes) have multiple possibilities to use the CoRR, after registration. There are available searching, subscribing and submitting tools, as well as instruments for downloading papers, adding references for the an entire journal or for a particular paper.

Each deposit into the CoRR is examined by reviewers in order to determine if the paper can be accepted for inclusion in the arXiv database. For each submitted paper, an average time of 24 hours is needed for analysis. After being accepted, a deposit receives a password used for later modification on its content and it can be viewed, searched and downloaded. The inclusion process will be explained later on.

The methodology associated with the submission process can be synthesized as in Figure 1.

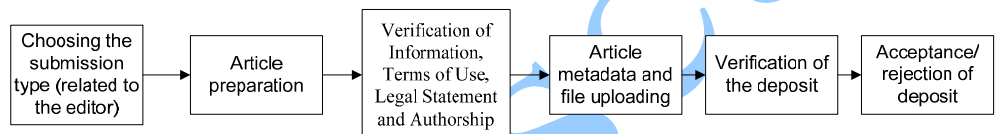

*Figure 1 Methodology for inclusion of an article in arXiv database*

The other specific database that was analyzed was the Directory of Open Journals (DOAJ) [Doa09], developed in Sweden at the Lund University Library. In order to include data to this archive, an on-line form has been completed, with information related to the journal (title, ISSN, publisher, a short description). The form is send to DOAJ Head Office which analyzed it by taking into account aspects related to the subject of the journal, the content, the language and information related to the regularity of appearance, the existence of an editorial board and the existence of a peer-review system for the selection of the articles. Also, the main requests were that the full text of the article has to be full text freely available and articles have to be published in Open Access system without an embargo period which means that the on-line publication of the paper is not conditioned by the publication in printed form (which can last a very long time). The response came after 3 weeks since the submission of the URL of the journal and it was a favorable one: "Annals. Computer Science Series" accomplishes all criteria established by DOAJ and it can be included in their database. The submission process is much simpler than for arXiv, metadata for DOAJ being created in a single step within the framework created by the Open Access Initiative.

Difference arise at this moment: while in arXiv the deposit consists in several articles (in PDF format) together with the metadata associated to each paper (included manually by some of the Editorial Board members), in DOAJ it is included the entire site of the publication. Metadata can be obtain here again, by completing an on-line form, not very different from that used by arXiv, which is producing specific metadata called DOAJ content.

The arXiv database does not index all published papers (not all papers can be associated with an arXiv category) and the situation for the three volumes considered for inclusion is presented in Figure 2 and each of the article is analyzed by a specialist in the field. Due to the metadata, articles can be found by introducing different keywords in the search box (the title, the name of the authors, reference of the conference, arXiv unique ID number, etc). An example is presented in Figure 3.

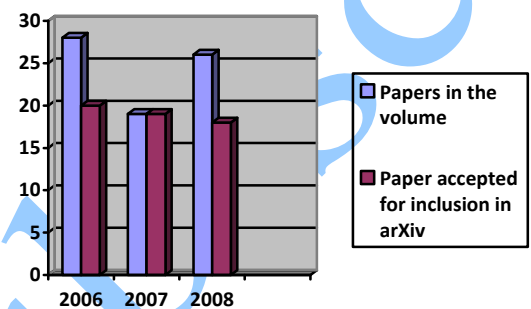

*Figure 2 Acceptance rate of articles in arXiv database*

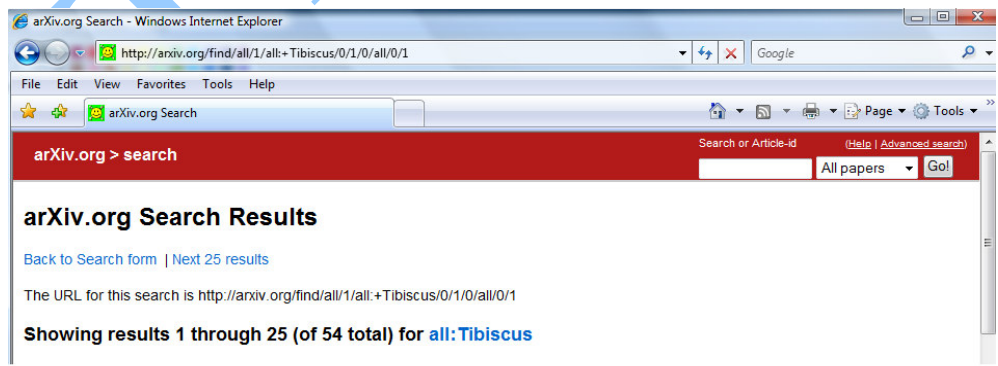

*Figure 3 Basic Search in arXiv database*

In the DOAJ database the information can be found in two ways:
- Direct search on the DOAJ site with the journal title, which returns the original site of the journal(http://anale-informatica.tibiscus.ro/?page=00_primapagina&lang=en);
- Click on the **DOAJ Content** icon which leads to the metadata section structured on the three existent volumes;
- A search on the **Find Articles** section after any of the fields that are included in the metadata. For example, in Figure 4 we present a search after an author's name.

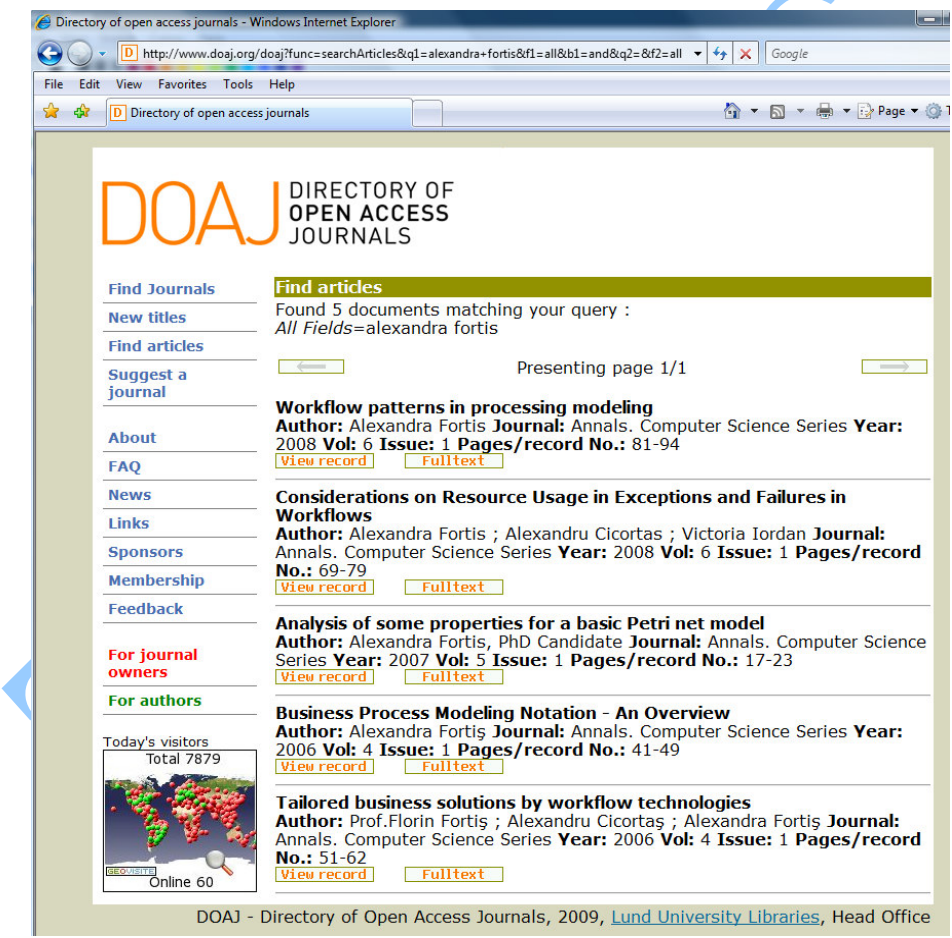

*Figure 4 Basic search on DOAJ database*

Through the inclusion of articles in those two databases the information (article metadata) has propagated towards other indexing systems, as follows:

137

- Data from DOAJ has spread through international library catalogs, as can be identifed in Figure 6, the search after the journal name on each catalog returnig as the result the original homepage of the journal;

- Data from arXiv was included in abstract databases as in Figure 5, from where it can be accesed at metadata level with information on authors, affiliation, paper status, abstract and acces to the full text stored in the arXiv database.

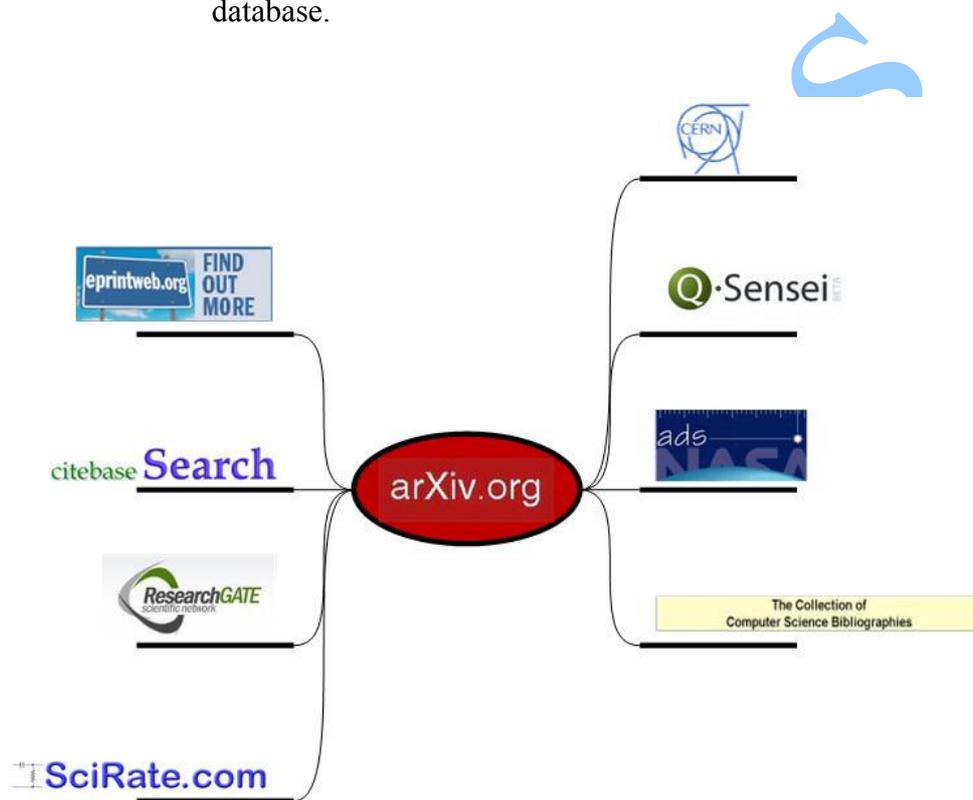

*Figure 5 Propagation of metadata from arXiv database*

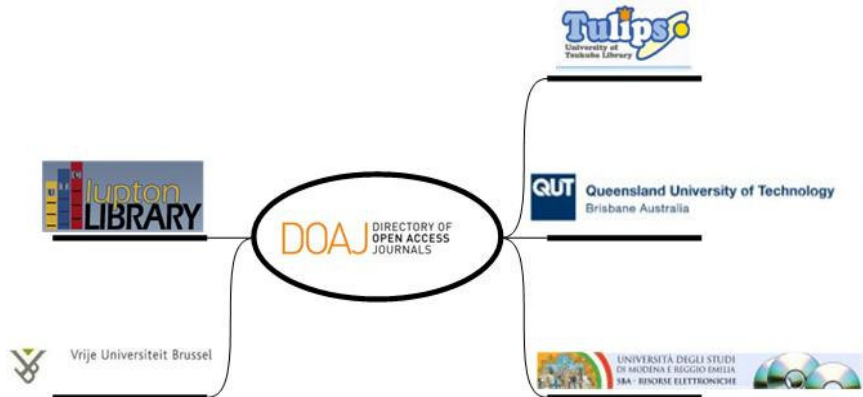

*Figure 6 Propagation of metadata from DOAJ database*

**Conclusions and future work**

In the moment of submission of this article, part of the research papers included in volumes from 2006 to 2008 were added into two different self-archiving databases (CogPrints, developed by University of Southampton, and Digital Library of the Commons (DLC), hosted by Indiana University). Some of the articles have already been included but the average time for the evaluation process is a month, so we could not perform a pertinent analysis related to the rate of inclusion and of their impact. The next stages in this action will be to adopt a management system for articles, according to OAI principles.

The Editorial Board intends to develop an archive containing metadata for all the editions published since 2003. Also, we will start the inclusion of scientific papers and abstracts in some open data-bases requiring laborious and time consuming activities (e.g. DLC – Digital Library of the Commons, Indiana University, USA, DBLP – Computer Science Bibliography, Trier University, Germany and CEEOL – Central and Eastern European Online Library, Frankfurt, Germany). For DBLP maintainers request a TOC entry for each paper submitted, which is equivalent to a bibliographic description of the paper, and it is created manually, for each paper.

In order to complete this task the editorial board will adopt the interoperability standards as mentioned in the **O**pen **A**ccess **I**nitiative and a protocol for metadata harvesting.

**References**

[arX09]   ArXiv e-print archive, Cornell University Library, http://arxiv.org

[Doa09]   Directory of Open Access Journals, http://www.doaj.org/

[Dub09]   The Dublin Core Metadata Initiative, http://dublincore.org/

[Pre05]   *Data Dictionary for Preservation Metadata*, OCLC Online Computer Library Center, Ohio, USA, 2005

[Fis08]   Julian. H. Fisher – *Scholarly Publishing Re-invented: Real Costs and Real Freedoms*, Ann Arbor, MI: Scholarly Publishing Office, University of Michigan, 2008

[Har02]   Stevan Harnard – *The Self-Archiving Initiative. Freeing the Refereed Reaserch Literature Online*, Southampton University, Highfield, UK, 2002

[HCB02]   Stevan Harnad, Les Carr, Tim Brody - *How and Why To Free All Refereed Research. From Access- and Impact-Barriers Online, No*w, Southampton University, Highfield, UK, 2002

[LS08]   Carl Lagoze, Herbert Van de Sompel - *The Open Archives Initiative Protocol for Metadata Harvesting*, 2008, available at http://www.openarchives.org/OAI/openarchivesprotocol.html

[Met09]   Metadata Encoding and Transmission Standard, Library of Congress, http://www.loc.gov/standards/mets/

[OAI02]   *Budapest Open Access Initiative*, 2002, available at http://www.soros.org/openaccess